\documentclass{ifacconf}

\usepackage{graphicx}      
\usepackage{natbib}        

\usepackage[subrefformat=parens]{subcaption}
\newcommand{\eqref}[1]{(\ref{#1})}

\newtheorem{assum}{Assumption}
\newtheorem{prop}{Proposition}
\newtheorem{prob}{Problem}
\newtheorem{defin}{Definition}

\begin{document}
\begin{frontmatter}

\title{Disconnection-aware Attack Detection in Networked Control Systems\thanksref{footnoteinfo}}


\thanks[footnoteinfo]{This work was supported in part by the Swedish Research Council (grant~2016-00861), the Swedish Foundation for Strategic Research (project CLAS), the Swedish Civil Contingencies Agency (project CERCES), and JST CREST Grant Number JPMJCR15K1, Japan.}

\author[KTH]{Hampei Sasahara} 
\author[TokyoTech]{Takayuki Ishizaki}
\author[TokyoTech]{Jun-ichi Imura}
\author[KTH]{Henrik Sandberg}

\address[KTH]{KTH Royal Institute of Technology,
   Stockholm, SE-100 44 Sweden (e-mail: \{hampei,hsan\}@kth.se).}
\address[TokyoTech]{Tokyo Institute of Technology,
   Tokyo, 1528552 Japan (e-mail: \{ishizaki,imura\}@sc.e.titech.ac.jp)}

\begin{abstract}                
This study deals with security issues in dynamical networked control systems.
The goal is to establish a unified framework of the attack detection stage, which includes the four processes of monitoring the system state, making a decision based on the monitored signal, disconnecting the corrupted subsystem, and operating the remaining system during restoration.
This paper, in particular, considers a disconnection-aware attack detector design problem.
Traditionally, observer-based attack detectors are designed based on the system model with a fixed network topology and cannot cope with a change of the topology caused by disconnection.
The disconnection-aware design problem is mathematically formulated and a solution is proposed in this paper.
A numerical example demonstrates the effectiveness of the proposed detector through an inverter-based voltage control system in a benchmark model.
\end{abstract}


\end{frontmatter}





\section{Introduction}
This study deals with security issues for physical networked systems, which face challenges in ensuring a safe and secure operation as in~\cite{Ding2018A}.
In modern and future cyber-physical systems, components are densely interconnected in both information and physical layers
and as a result once an adversarial attack irrupts its effects propagate over a broad range in the entire networked system.
In fact, several incidents have been reported and it covers a wide range of applications including an attack on an uranium enrichment plant, attacks on Ukrainian power plants, remote car hacking, and unmanned aerial vehicle hacking, see~\cite{Kushner2013Real,Lee2016Analysis,Charlie2015Remote,Daniel2012Drone}.

For secure operation of physical systems, novel security techniques in the physical layer are required aside from the existing information security techniques because of difference between the requirements of information systems and physical systems.
For instance, patching and frequent updates are not well suited for control systems as mentioned in~\cite{Alvaro2011Attacks}.
Furthermore, implementing security technologies in physical layers in addition to information layers fits the notion of ``defense in depth'' advocated in~\cite{Kulpers2006Control}, which argues the importance of multiplication of protections.

When considering security in the physical layer we have to handle dynamics of the system because in most cases inertia, or several counterparts, cannot be ignored in physical systems. 
In~\cite{Seyed2019Systems}, a survey paper on security for dynamical systems from the control theoretical perspective, the defense mechanism of physical systems against attacks is classified into three stages: prevention, resilience, and detection stages.
The process of the detection stage is further divided into the four steps: monitoring, decision, disconnection, and restoration steps.
In the monitoring step, the state of the operation is observed.
Based on the monitored signal, decision on attack injection is made in the decision step.
Then the attacked components are disconnected for suppressing propagation of attack effects.
Finally, after removing the cause of the attack the isolated part is reconnected.
Traditionally, those steps have separately been studied based on the assumption that the steps are able to operate independently.

However, those steps should be discussed in a unified manner especially for dynamical networked systems under control, because unplugging some input ports would lead to loss of the original function owing to disconnection of feedback in controlled systems.
For example, when an attack is injected into one of the reference signals, simply disconnecting the input port of the reference signal results in deviation of all output signals from the reference.
In the worst case, stability of the system would be lost if the signal under attack configures a closed-loop system.
Therefore, it is required to design the network system with a protection system to be compatible with disconnection and to discuss the entire security flow in the detection stage in an integrated fashion.

The goal of this study is to establish a unified framework of the detection stage.
First, the system must be resilient to disconnection of a part of the networked system.
In other words, the system is needed to be able to continually operate even when disconnections happen as a reaction against attacks.
Since existing researches (e.g.,~\cite{Ishizaki2018Retrofit,Anderson2019System,Zhihua2014Modularized,Akhavan2019Passivity}) enable design of systems resilient to disconnection, in this paper we suppose that the entire system can be constructed to have the resilience property to disconnection.
Based on this premise, this paper treats the design problem of a distributed attack detector compatible with disconnection.
Conventionally, observer-based residual generators, which monitor the state of the system operation by comparing the ideal signal under the normal operation and the actual measured signal, are designed based on the system model with a fixed network topology.
The design policy leads to lack of the capability to handle topology changes caused by disconnection.
We mathematically formulate the disconnection-aware dynamical attack detector design problem and propose a solution based on \emph{retrofit control} developed in~\cite{Ishizaki2018Retrofit,Ishizaki2019Modularity}.
The effectiveness of the proposed approach is illustrated through a numerical example of power distribution network systems.

Several related works that propose design methods of physical systems being secure throughout the entire detection stage can be found.
In~\cite{Sasaki2015Model,Sasaki2017Model}, the authors have proposed fallback control.
In the framework, fallback operation is switched to as an incident response of the control system where facilities are preferentially protected and minimal system that can continue to operate during cyber attacks.
Although the attack detection process in fallback control is achieved through a switched Lyapunov function, it is difficult to find a switched Lyapunov function for large-scale systems.
A system design method that can perform plug-and-play fault detection and control-reconfiguration in a unified manner is proposed in~\cite{Riverso2016Plug}, the method of which can possibly be applied to attack detection as well.
Although the existing study considers large-scale systems and the proposed method is scalable, it requires a technical assumption that interconnection signals between subsystems are constrained in bounded sets.
The assumption means that each subsystem can be regarded as a decoupled system with bounded disturbances and is not well suited for stability analysis of strongly connected networked systems.

This paper is organized as follows.
In Sec.~\ref{sec:uni}, we first review security operation for dynamical networked systems.
Then we provide a motivating example of power distribution network systems to highlight the necessity of a unified framework of the detection stage.
Moreover, the disconnection-aware distributed attack detector design problem is mathematically formulated.
We review retrofit control, which is the key to solve the formulated problem in Sec.~\ref{sec:prop},
and subsequently, we provide a solution to the formulated problem based on retrofit control.
Sec.~\ref{sec:num} provides a numerical example that illustrates the effectiveness of the proposed method and Sec.~\ref{sec:conc} draws conclusion.

{\bf Notation}
We define a shorthand of a transfer function $G(s)=C(sI-A)^{-1}B+D$ by
\[
 \left[
 \begin{array}{c|c}
 A & B \\ \hline
 C & D
 \end{array}
 \right]:= G(s).
\]

\section{Unified Framework for Attack Detection and Problem Formulation}
\label{sec:uni}

In this section, we review the security flow for dynamical networked systems under attack through their mathematical models.
Then we provide a motivating example highlighting the necessity of a unified framework of the detection stage.
Supposing that the system is resilient to disconnection, we formulate the problem of distributed attack detector design in which the distributed attack detector can cope with topology variation caused by disconnection.

\subsection{Security Operation for Dynamical Networked Systems}
\label{subsec:model}

Consider a linear time-invariant interconnected system $\Sigma$ composed of $N$ subsystems
\begin{equation}\label{eq:sysi}
 \Sigma_i:
 \left\{
 \begin{array}{cl}
 \dot{x}_i & = \textstyle{A_i x_i + L_iv_i+B_i r_i+a_{i,x}}\\
 w_{i} & = \textstyle{W_ix_i+Z_iv_i+U_i r_i + a_{i,w}}\\
 y_{i} & = \textstyle{C_ix_i+E_iv_i + D_i r_i} + a_{i,y}
 \end{array}
 \right.
\end{equation}
for $i=1,\ldots,N$ with an interconnection $v_i = \sum_{j \in \mathcal{N}_i}M_{ij}w_j$,
where $x_i,v_i,w_i,y_i,r_i$ denote the state, the interconnection input, the interconnection output, the measurement output, and the reference signal, respectively.
The index set associated with the neighborhood of $\Sigma_i$ is denoted by $\mathcal{N}_i$.
The exogenous inputs $a_{i,x},a_{i,w},a_{i,y}$ represent the effects to the system behavior caused by the attack.
For simplicity, the entire interconnected system is assumed to be well-posed throughout this paper.

\emph{Example:}
The system model can represent a wide range of attacks, such as controller hijacking executed by Stuxnet, see~\cite{Nicolas2011Stuxnet}.
Let us suppose that the $i$th subsystem dynamics without attacks is represented as $\dot{x}_i = A_ix_i+L_iv_i+B_i r_i+u_i$ where $u_i$ is a control input generated by a feedback controller implemented through programmable logic controllers (PLCs).
Stuxnet overwrites the control algorithm in PLCs and the resulting control input $u_i$ becomes a malicious input $\alpha_i$, whose effects to the system behavior is modeled by~\eqref{eq:sysi} with $a_{i,x}:=\alpha_i-u_i$.

Defense mechanisms of physical systems can be categorized into three stages: prevention, resilience, and detection as in~\cite{Seyed2019Systems}.
In this study, we pay attention to the detection stage, the process of which includes monitoring the state of the system operation for detection of attacks, making a decision of attack injection, disconnecting the part under the attack for suppressing propagation of the influence of the attack, and restoring the normal operation.
The process of each step, which exploits techniques in fault detection and isolation (\cite{Ding2013Model}) is reviewed in the following.

As the first step of the detection stage, residual generation is performed permanently during the operation.
The residual $\epsilon (t)$ is generated by comparing the ideal output signal $\hat{y}(t)$ under the normal operation and the actual measurement output $y(t)$ by $\epsilon (t) := y(t)-\hat{y}(t)$, which is normally zero when no attack is injected if measurement noises can be ignored.
The ideal output signal $\hat{y}(t)$ is generated through an observer with the information on the input signals and the dynamical system model during normal operation.
For instance, for the interconnected system composed of~\eqref{eq:sysi}, a distributed observer, each of whose subobservers interacts an estimated interconnection signal $\hat{w}_i$ to each other, can be built as
\begin{equation}\label{eq:obnomi}
 \hat{\Sigma}_i:
 \left\{
 \begin{array}{cl}
 \hat{x}_i & = \textstyle{A_i\hat{x}_i + L_i\hat{v}_i + B_i r_i+H_i(y_i-\hat{y}_i)}\\
 \hat{w}_i & = \textstyle{W_i\hat{x}_i + Z_i\hat{v}_i +U_i r_i}\\
 \hat{y}_i & = \textstyle{C_i\hat{x}_i+E_i\hat{v}_i+ D_i r_i}\\
 \hat{v}_i & = \textstyle{\sum_{j \in \mathcal{N}_i} M_{ij} \hat{w}_j}
 \end{array}
 \right.
\end{equation}
with appropriate observer gains $H_i$.
At the next step, based on the instantaneous value of the residual $\epsilon(t)$, decisions are made by a testing method, such as the chi-squared test (\cite{Mehra1971An}), the generalized likelihood ratio test (\cite{Willsky1976A}), and the sequential hypothesis test (\cite{Alvaro2011Attacks}).
When a decision of existence of attack is made, isolation, namely, identifying the location of the attack~(\cite{Pasqualetti2015Control}) is performed as well.
Then for eliminating influence of the attack the input port is isolated by disconnecting the facilities under attack, e.g., turning off the switch of the interaction between the plant and the controller.
Finally, the system is restored to regain its original function by removing the source of the attack and reconnecting the islanded components.

\subsection{Motivating Example}

\begin{figure}[h]
\begin{center}
\includegraphics[width=.98\linewidth]{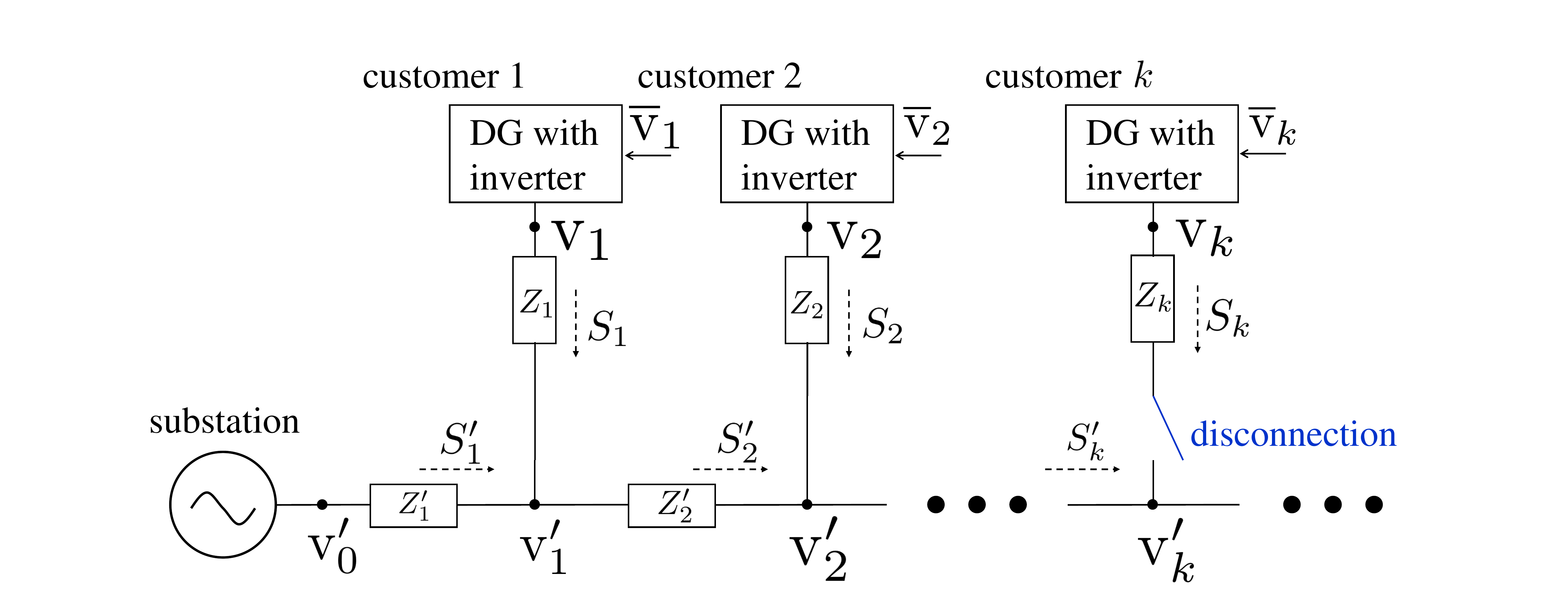}    
\caption{Infrastructure of a low-voltage distribution network system.}
\label{fig:DN}
\end{center}
\end{figure}

This subsection provides a motivating example that demonstrates operation failures owing to disconnection of the components under attack.
We consider low-voltage power distribution networks as shown in Fig.~\ref{fig:DN}.
In the distribution network system, each customer has a distributed generation (DG) with which an inverter is equipped as a controller.
The purpose of the control is to regulate the voltage magnitudes ${\rm v}_k$ to given reference values $\overline{{\rm v}}_k$, which is identically set to be 230 [V] (for the detail of the modeling, see Sec.~\ref{sec:num}).
As a residual generator, we design a distributed observer~\eqref{eq:obnomi} with certain observer gains $H_i$ designed to guarantee stability of the error dynamics for the original system $\Sigma$.

\begin{figure}[h]
\begin{center}
\includegraphics[width=.98\linewidth]{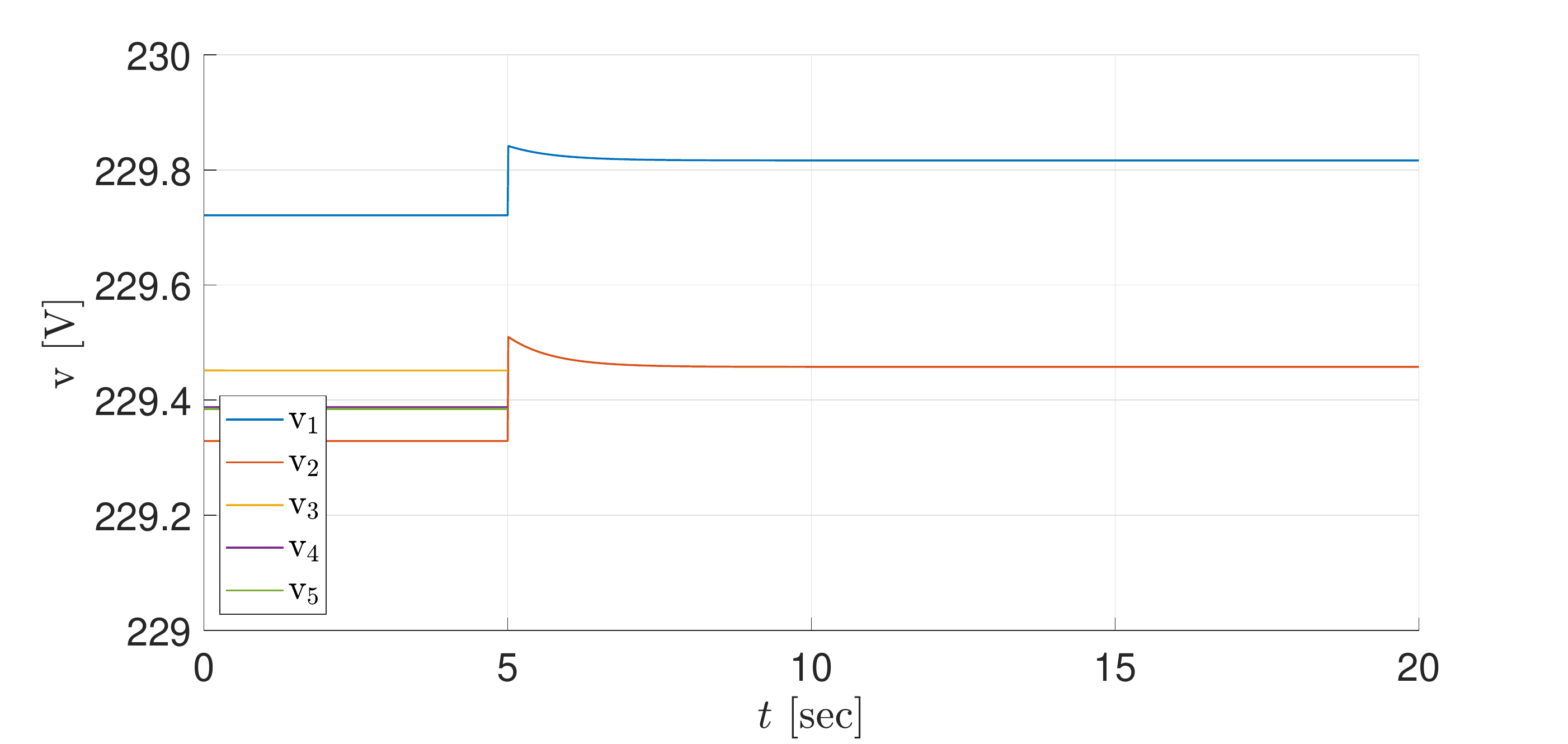}
\caption{Time series of the voltage magnitudes under disconnection at $t_0=5$.}
\label{fig:v_discon}
\end{center}
\end{figure}

\begin{figure}[h]
\begin{center}
\includegraphics[width=.98\linewidth]{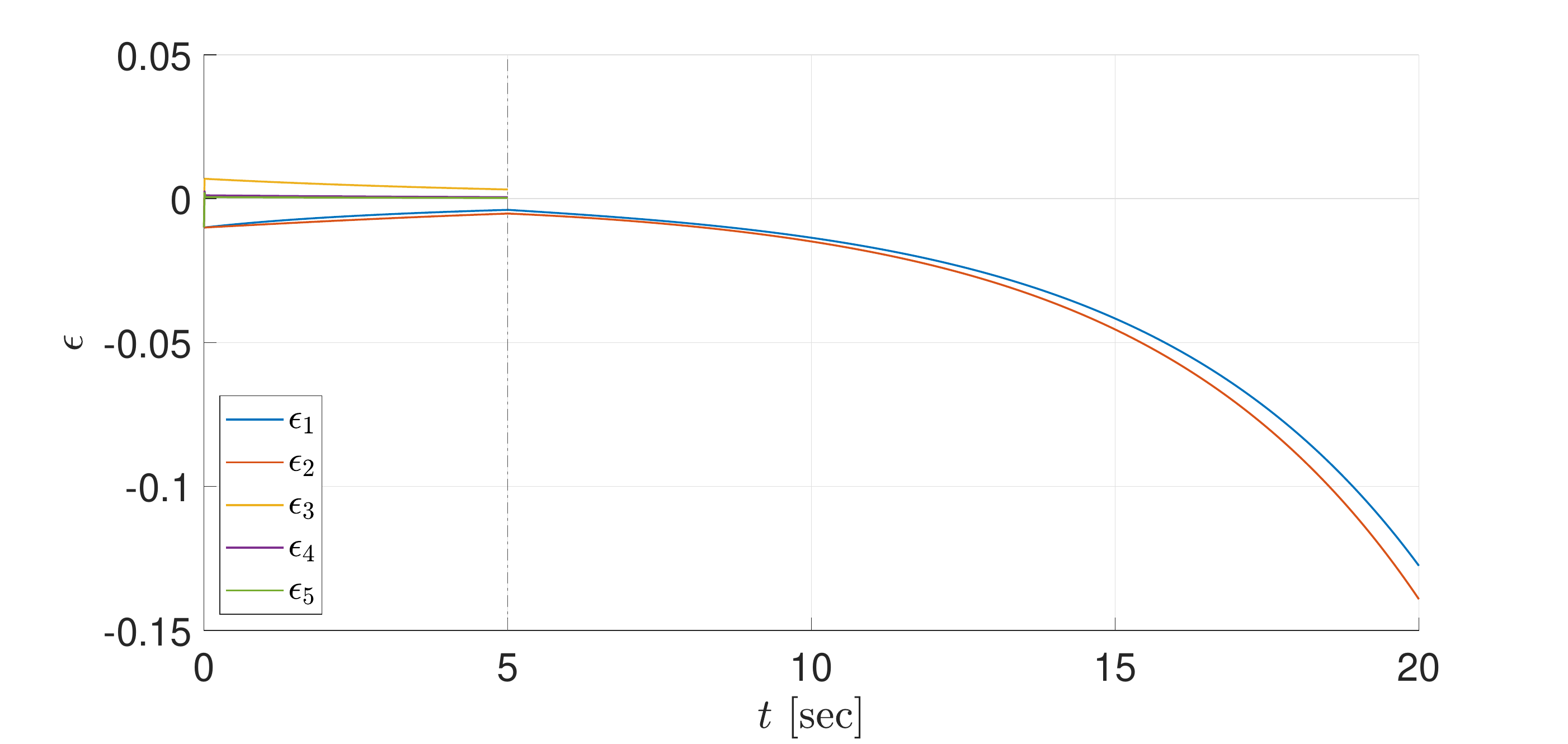}
\caption{Time series of the residual with a distributed observer~\eqref{eq:obnomi} under disconnection at $t_0=5$.}
\label{fig:example}
\end{center}
\end{figure}

Let us observe behaviors of the system under some disconnections caused by attacks.
The distribution network has five customers and we suppose that an attack is injected into the $i$th DGs for $i=3,4,5$.
As a reaction against the attack, it is supposed that the DGs under the attack are disconnected at the time $t_0=5$.
Fig.~\ref{fig:v_discon} illustrates the time series of the voltage magnitudes under the disconnection at $t_0=5$.
It can be confirmed that the voltage magnitudes in the remaining parts are kept to be around the setpoint even after the disconnection.
This result implies that the distribution network system can work well under the disconnection and is resilient to disconnection caused by attacks.
On the other hand, Fig.~\ref{fig:example} illustrates the time series of the residual with a distributed observer~\eqref{eq:obnomi} under disconnection at $t_0=5$, which is represented by the dash-dot line.
It can be observed that, although the estimation error converges to zero before the disconnection, the error diverges after the disconnection.
The instability is arisen from the inadequate choice of the observer gains $H_i$, which are designed only for the original system without disconnection.
The resulting distributed observer is not compatible with network topology change caused by the disconnection.

As shown in the example, the conventional attack detector would lead to loss of monitoring function owing to instability of the distributed observer when disconnection happens.
Therefore, it is required to develop a novel attack detector design method that can cope with disconnection.
Moreover, since the system dynamics is changed by the disconnection, all processes that utilize system model information have to be designed taking the system variation into account explicitly.
Based on the observation, we consider developing a unified framework for the detection stage in the next subsection.

\subsection{Disconnection-aware Attack Detector Design Problem}
\label{subsec:prob}


The motivating example above has shown the necessity of a novel framework for the security flow:
\begin{enumerate}
\setcounter{enumi}{-1}
\item design process: designing the interconnected system $\Sigma$ to be resilient to disconnection caused by attacks,
\item monitoring process: monitoring the local time series of the residual $\epsilon_i(t)$ generated by comparing the actual measurement signal $y_i(t)$ with the ideal signals $\hat{y}_i(t)$ created through a distributed observer,
\item decision process: making a decision of attack and the attacked place at each detector based on the local monitored residual $\epsilon_i(t)$,
\item disconnection process: disconnecting the $i$th subsystem where an attack is detected, namely, $w_i$ is set to be zero,
\item operation and restoration process: operating the remaining interconnected system during restoration.
\end{enumerate}
Clearly, at least the first three processes from the top of the list are required to be compatible with disconnection.
Our goal is to establish a unified framework for the entire security flow.

In this study, we regard stability under disconnection as resiliency of the networked system.
We treat the resiliency of the original system as a premise.
The following assumption is made.
\begin{assum}\label{assum:res}
For any $\mathcal{I} \subset \{1,\ldots,N\},$ the remaining interconnected system $\Sigma_{\mathcal{I}}$ is internally stable where $\Sigma_{\mathcal{I}}$ is given by the interconnected system composed of the subsystems $\Sigma_i$ for $i\in \mathcal{I}$ with $w_j=0$ for $j \notin \mathcal{I}$ in~\eqref{eq:sysi}.
\end{assum}
Assumption~\ref{assum:res} states that the stability of the interconnected system itself is guaranteed for any disconnection with respect to the interconnection between the subsystems.
Although the detail of designing the interconnected system satisfying Assumption~\ref{assum:res} is not focused on in this paper,
we provide brief description of two possible approaches for the design method in the following.
The first approach is given by simply applying existing methods in the design process for designing a dynamical interconnected system to be resilient to disconnection, see~\cite{Ishizaki2018Retrofit,Anderson2019System,Zhihua2014Modularized,Akhavan2019Passivity}.
In this approach the points to be disconnected are naturally determined and subsystems $\Sigma_i$ are given as the designed components.
In the other approach, for a given system $\Sigma$, we choose the interconnection points to be disconnected such that Assumption~\ref{assum:res} is satisfied.
The pick of connecting points leads to the system description as an interconnected system of $\Sigma$ composed of the subsystems $\Sigma_i$.
The policy of deciding interconnection points can specifically be given when the entire system is built by a negative feedback interconnection of passive systems (\cite{Bernard2006Dissipative}).
In this case, Assumption~\ref{assum:res} is satisfied simply choosing each element as a subsystem.
If the original system does not have such a property, connecting points can be determined through off-line simulation in advance.

In this paper, we focus on the problem of designing a distributed observer to generate $\hat{y}$ in the monitoring process as a preliminary step.
The difficulty here is that, even if the dynamics of the error signal $e:=x-\hat{x}$ with the distributed observer in the form of~\eqref{eq:obnomi} with fixed observer gains $H_i$ is stable,
the remaining error signal dynamics under some disconnection is possibly destabilized with the original observer gains, which are compatible only with the original fully interconnected system $\Sigma$.

We formulate the dynamical attack detector design problem as follows.

\begin{prob}\label{prob}
Under Assumption~\ref{assum:res}, design a distributed observer
\[
 (\hat{x}_i,\hat{y}_i,\hat{w}_i) = \mathcal{O}_i(y_i,\{\hat{w}_j\}_{j \in \mathcal{N}_i},r_i)
\]
such that $\hat{x}_{\mathcal{I}} \to x_{\mathcal{I}}$ and $\hat{y}_{\mathcal{I}} \to y_{\mathcal{I}}$ as $t \to +\infty$ for any $\mathcal{I} \in \{1,\ldots,N\}$ under any initial conditions when the attack signals are zero,
where $\hat{x}_{\mathcal{I}}, \hat{y}_{\mathcal{I}}$ are the states and outputs of $\mathcal{O}_{\mathcal{I}}$, which is given by the distributed observers composed of $\mathcal{O}_i$ for $i\in \mathcal{I}$ with $\hat{w}_j=0$ for $j \notin \mathcal{I}$.
\end{prob}

The implicit assumptions on the problem are as follows:
\begin{itemize}
\item The network topology of the distributed observer is the same as that of the original interconnected system and
\item the input signals $\hat{w}_j$ and $r_i$ are not fabricated by the attacker.
\end{itemize}

Note that a naive approach to guarantee stability under network topology variation caused by disconnection is to not utilize feedback of the error signal, namely, setting $H_i=0$ for any $i$ in~\eqref{eq:obnomi}.
Then the stability of the error signal dynamics is preserved owing to Assumption~\ref{assum:res} because it is equivalent to the stability of the original system.
However, the poles of the error signal dynamics cannot be shifted without feedback and hence early attack detection cannot be performed with the naive approach.
Note also that, if we represent the stability condition as linear matrix inequalities (LMIs),
the number of the constraints reaches $2^N$ and hence it is practically impossible to solve the problem via the naive LMI approach.

\emph{Remark:}
White-box attacks (\cite{Anirban2018Adversarial}), namely, attacks with full knowledge about the system architecture are out of focus at the present stage.
For example, if the attacker can arbitrarily choose the attack signals $a_{i,x}$ and $a_{i,y}$, it can be readily performed to generate an attack signal that cannot be detected by any possible detectors.
Furthermore, even without $a_{i,y}$ if there exists zero-dynamics in the dynamical system undetectable attacks can be constructed as shown in~\cite{Pasqualetti2015Control}.
Since these attacks cannot be coped with by detector design alone,
such problems are not treated here.

\section{Proposed Disconnection-aware Attack Detector Design Method}
\label{sec:prop}

In this section, we propose a design method of a distributed observer that meets the requirement in the formulated problem.
First, we briefly review retrofit control, developed in~\cite{Ishizaki2018Retrofit,Ishizaki2019Modularity}, which plays the key role to solve the problem.
Subsequently, we propose a solution based on the retrofit control.

\subsection{Brief Review of Retrofit Control}
\label{subsec:retro}

Retrofit control is a newly developed controller design method the main purpose of which is to achieve modular design of large-scale interconnected control systems.
In the framework, each subcontroller is supposed to be designed only using the model information of the subsystem to which the subcontroller is attached.
For the controller design it is naturally required that the function held by the entire interconnected system must be preserved even with introduction of subcontrollers in a sequential manner.
Retrofit control provides a specific controller design method that fulfills this requirement.

\begin{figure}[h]
\centering
\begin{tabular}{cc}
  \begin{minipage}[t]{.46\linewidth}
    \centering
    \includegraphics[width=.98\linewidth]{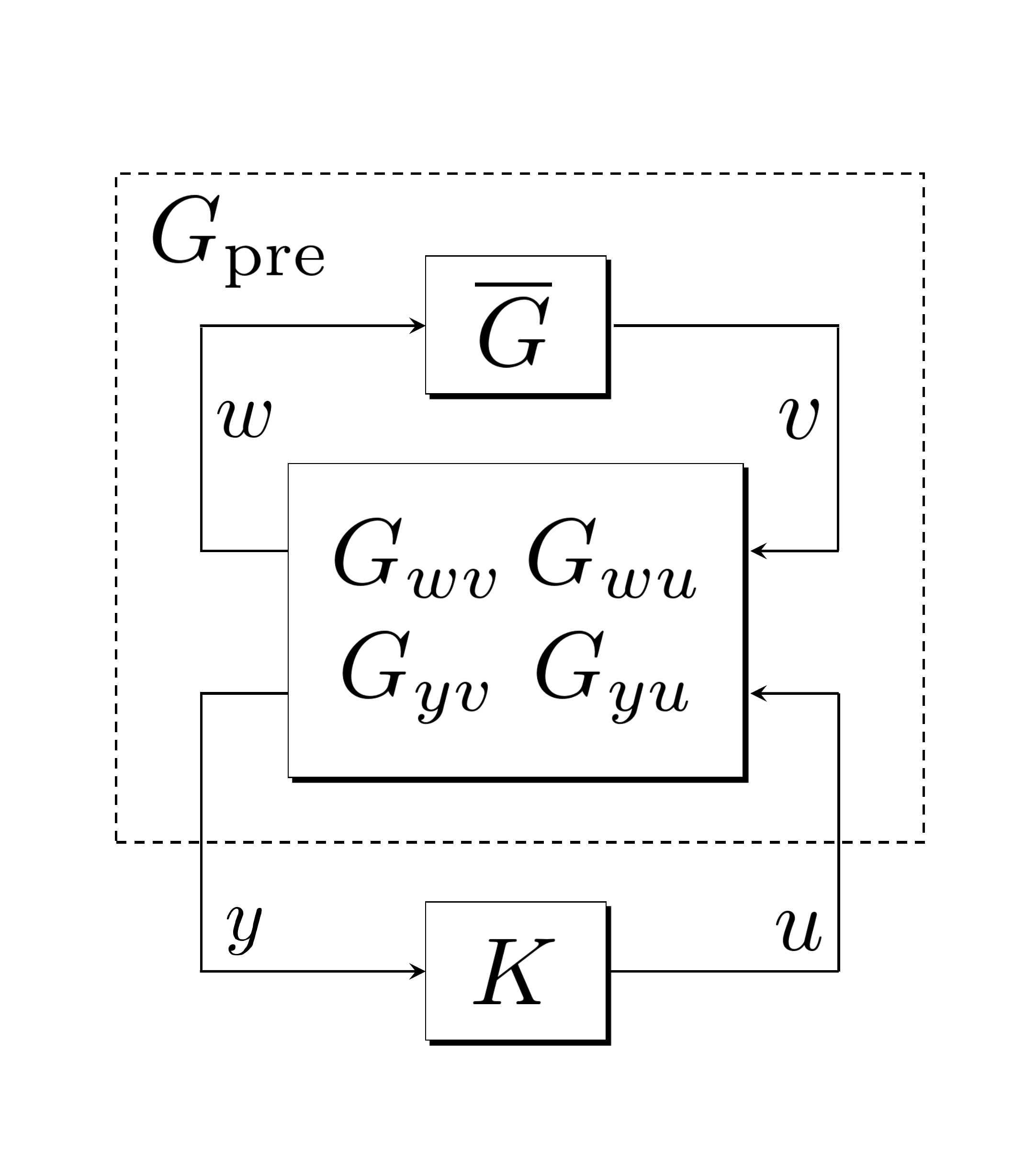}
    \subcaption{Entire interconnected system to be controlled in the retrofit control framework.}
  \end{minipage} &
  \begin{minipage}[t]{.46\linewidth}
    \centering
    \includegraphics[width=.98\linewidth]{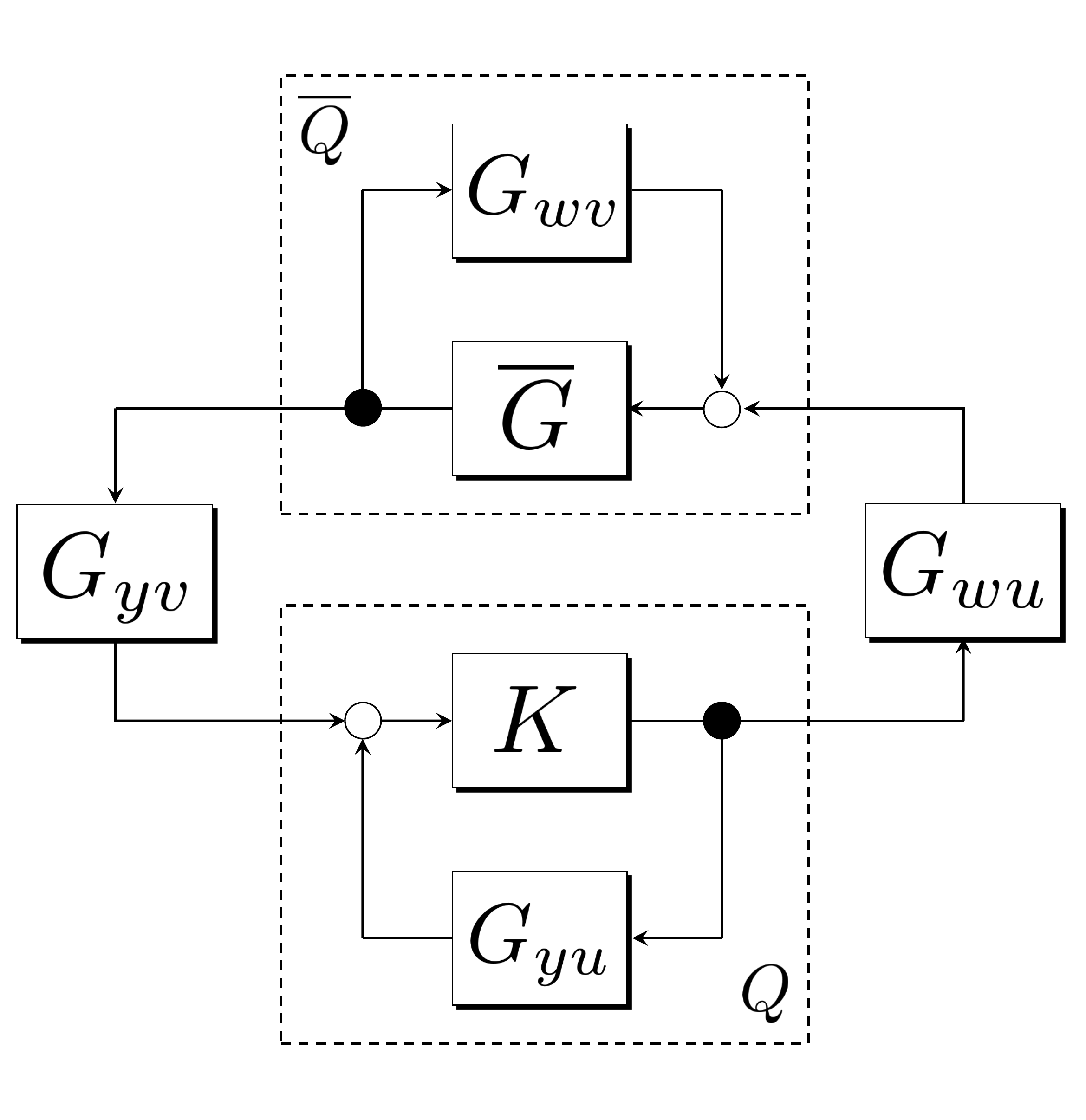}
    \subcaption{Equivalent system explicitly representing the Youla parameters $Q$ and $\overline{Q}$.}
  \end{minipage}
\end{tabular}
  \caption{Equivalent block diagrams of the closed-loop system.}
  \label{fig:RetroSchema}
\end{figure}

The schematic diagram of the interconnected system to be controlled in the retrofit control framework is depicted in Fig.~\ref{fig:RetroSchema}~(a).
It is assumed that the model information of $G=(G_{wv},G_{wu},G_{yv},G_{yu})$, the subsystem associated with the subcontroller to be designed, is available while the model information of $\overline{G}$, the other part of the interconnected system, is unavailable.
We suppose that the interconnected system without the controller $K$, denoted by $G_{\rm pre}$, stably operates and consider attaching $K$ to improve the control performance while preserving the stability.
To mathematically handle the problem, we first consider the following set
\[
 \overline{\mathcal{G}} := \{\overline{G}: G_{\rm pre}\mathrm{\ is\ internally\ stable.}\}
\]
and provide the definition of retrofit controllers as follows.
\begin{defin}
 The controller $K$ in Fig.~\ref{fig:RetroSchema}~(a) is said to be a retrofit controller if the entire feedback system in Fig.~\ref{fig:RetroSchema}~(a) is internally stable for any $\overline{G} \in \overline{\mathcal{G}}$.
\end{defin}

A necessary and sufficient condition for retrofit controllers is given as follows:
\begin{prop}\label{fact:retro}
Assume that $G$ in Fig.~\ref{fig:RetroSchema}~(a) is stable.
Then $K$ is a retrofit controller if and only if $G_{wu}QG_{yv}=0$ and $Q$ is stable where $Q:=K(I-G_{yu}K)^{-1}$ denotes the Youla parameter with respect to $K$ for $G_{yu}$.
\end{prop}
The interpretation of the conditions is given as follows.
Fig.~\ref{fig:RetroSchema}~(b) is obtained by transforming the system Fig.~\ref{fig:RetroSchema}~(a) with the Youla parameters $Q$ and $\overline{Q}$, which correspond to $K$ and $\overline{G}$, respectively.
Since $\overline{Q}$ is an arbitrary stable transfer matrix, the condition is necessary and sufficient for internal stability.
For the complete proof, see~\cite{Ishizaki2019Modularity}.

\subsection{Proposed Design Method}
\label{subsec:prop}

We propose a disconnection-aware distributed attack detector design method as a solution to Problem~\ref{prob} based on retrofit control.
Let us consider a distributed observer in the following form:
\begin{equation}\label{eq:dob}
 \mathcal{O}_i:
 \left\{
 \begin{array}{cl}
 \dot{\hat{x}}_i & = \textstyle{A_i\hat{x}_i+L_i\sum_{j \in \mathcal{N}_i}M_{ij}\hat{w}_j + B_i r_i + \mu_i}\\
 \hat{w}_i & = \textstyle{W_i \hat{x}_i + Z_i\sum_{j \in \mathcal{N}_i}M_{ij}\hat{w}_j + U_ir_i + \nu_i}\\
 \hat{y}_i & = \textstyle{C_i\hat{x}_i + E_i\sum_{j \in \mathcal{N}_i}M_{ij}\hat{w}_j + D_ir_i}
 \end{array}
 \right.
\end{equation}
where $\mu_i$ and $\nu_i$ are artificial signals injected into the subobserver.
It should be noted that the distributed observer has the form of~\eqref{eq:obnomi} if we choose $\mu_i=H_i(y_i-\hat{y}_i)$ and $\nu_i=0$.

Let $e_i:=\hat{x}_i-x_i$ be the local state estimation error.
Then the dynamics of $e_i$ is represented as
\[
 \mathcal{E}_i:
 \left\{
 \begin{array}{cl}
 \dot{e}_i & = \textstyle{ A_i e_i + L_i\sum_{j \in \mathcal{N}_i}M_{ij}\omega_j + \mu_i}\\
 \omega_i & = \textstyle{ W_i e_i + Z_i\sum_{j \in \mathcal{N}_i}M_{ij}\omega_j + \nu_i}\\
 \psi_i & = \textstyle{ C_i e_i + E_i\sum_{j \in \mathcal{N}_i}M_{ij}\omega_j}
 \end{array}
 \right.
\]
where $\omega_i:=\hat{w}_i-w_i$ and $\psi_i:= \hat{y}_i-y_i$ denote the local estimation errors in terms of the interconnection signal and the measurement output, respectively.
Obviously, Problem~\ref{prob} is equivalent to the stabilization problem of the interconnected systems composed of $\mathcal{E}_i$ under disconnection as long as the distributed observer is assumed to have the form in~\eqref{eq:dob}.
Therefore, we consider designing a stabilizing controller $K_i: \psi_i \mapsto (\mu_i,\nu_i)$ in the following discussion.

\begin{figure}[h]
\begin{center}
\includegraphics[width=.98\linewidth]{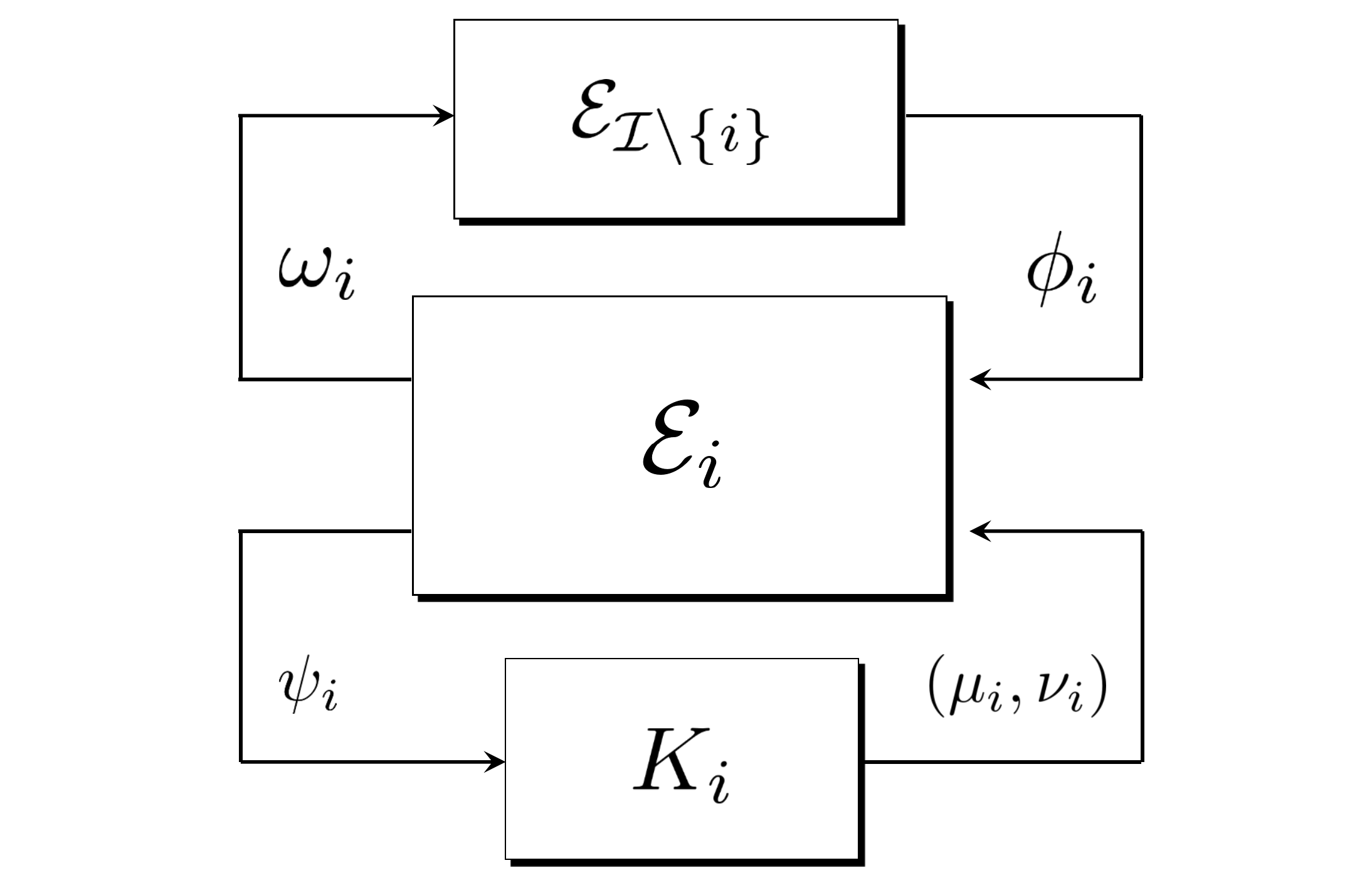}
\caption{Block diagram of $\mathcal{E}_{\mathcal{I}}$ for a particular $\mathcal{I}$ from the perspective of $\mathcal{E}_i$ where $\phi_i:= \sum_{j \in \mathcal{N}_i}M_{ij}\omega_j$.}
\label{fig:block}
\end{center}
\end{figure}

Now we take a particular $\mathcal{I} \subset \{1,\ldots,N\}$ and let $\mathcal{E}_{\mathcal{I}}$ be the interconnected system composed of $\mathcal{E}_i$ for $i\in \mathcal{I}$.
Then the block diagram of $\mathcal{E}_\mathcal{I}$ from the perspective of the $i$th subsystem $\mathcal{E}_i$ can be represented by Fig.~\ref{fig:block} where $\mathcal{E}_{\mathcal{I}\setminus \{i\}}$ is the interconnected system composed of $\mathcal{E}_j$ for $j \in \mathcal{I}\setminus \{i\}$ and $\phi_i := \sum_{j\in \mathcal{N}_i}M_{ij}\omega_j$.
From Assumption~\ref{assum:res} and Proposition~\ref{fact:retro}, if $K_i$ satisfies the condition for retrofit controllers for any $i \in \{1,\ldots,N\}$, then the resulting distributed observer with the controllers $K_i$ meets the requirement of the problem.

The difficulty here is that the procedure to design a retrofit controller for $K_i$ is not obvious.
The following theorem, the main result of this paper from a theoretical perspective, provides a retrofit controller design method for $K_i$.
\begin{thm}
Design $K_i$ as
\begin{equation}\label{eq:Ki}
 K_i: \left\{
 \begin{array}{cl}
 \dot{\chi}_i & = A_i \chi_i + \mu_i\\
 \mu_i & = H_i\psi_i\\
 \nu_i & = -W_i\chi_i
 \end{array}
 \right.
\end{equation}
with an observer gain $H_i$ such that $A_i+H_iC_i$ is Hurwitz.
Then $K_i$ satisfies the condition for retrofit control.
\end{thm}

\begin{pf}
Define $Q_i := K_i(I-G_{i, \psi  (\mu,\nu)}K_i)^{-1}$ with
\[
 G_{i, \psi  (\mu,\nu)} := \left[
 \begin{array}{c|c}
 A_i & [I\ 0]\\ \hline
 C_i & [0\ 0]
 \end{array}
 \right].
\]
First, as long as $Q_i$ is stable, the condition for retrofit control in Fig.~\ref{fig:block} is represented by $G_{i,\omega (\mu,\nu)} Q_i G_{i, \psi \phi}=0$ where
\[
 G_{i,\omega (\mu,\nu)} :=
 \left[
 \begin{array}{c|c}
 A_i & [I\ 0]\\ \hline
 W_i & [0\ I]
 \end{array}
 \right],\quad G_{i, \psi \phi}:=
 \left[
 \begin{array}{c|c}
 A_i & L_i\\ \hline
 C_i & E_i
 \end{array}
 \right].
\]
Clearly,
\begin{equation}\label{eq:GK}
 G_{i,\omega (\mu,\nu)}K_i=0
\end{equation}
is a sufficient condition.
We show that~\eqref{eq:Ki} satisfies~\eqref{eq:GK} and also that $Q_i$ is stable.

The transfer matrix $G_{i,\omega (\mu,\nu)}$ can be rewritten by
\[
 G_{i,\omega (\mu,\nu)} = \left[
 \left[
 \begin{array}{c|c}
 A_i & I\\ \hline
 W_i & 0
 \end{array}
 \right]\ I\right].
\]
and the transfer matrix of $K_i$ is given by
\[
 \begin{array}{ccc}
 K_i & =
 \left[
 \begin{array}{c|c}
 
 A_i & H_i\\ \hline
 \left[
 \begin{array}{c}
  0\\
 -W_i
 \end{array}\right] & \left[
 \begin{array}{c}
 H_i\\
 0
 \end{array}\right]
 \end{array}
 \right]
 & =\left[
 \begin{array}{c}
 I\\
 \left[
 \begin{array}{c|c}
 A_i & I\\ \hline
 -W_i & 0
 \end{array}
 \right]
 \end{array}\right] H_i.
 \end{array}
\]
Thus, we have~\eqref{eq:GK}.
Moreover, since
\[
 G_{i,\psi (\mu,\nu)}K_i = \left[
 \begin{array}{c|c}
 A_i & H_i\\ \hline
 C_i & 0
 \end{array}
 \right],
\]
if $A_i+H_i C_i$ is Hurwitz then $Q_i$ is stable.
\qed
\end{pf}

\begin{figure}[h]
\begin{center}
\includegraphics[width=.98\linewidth]{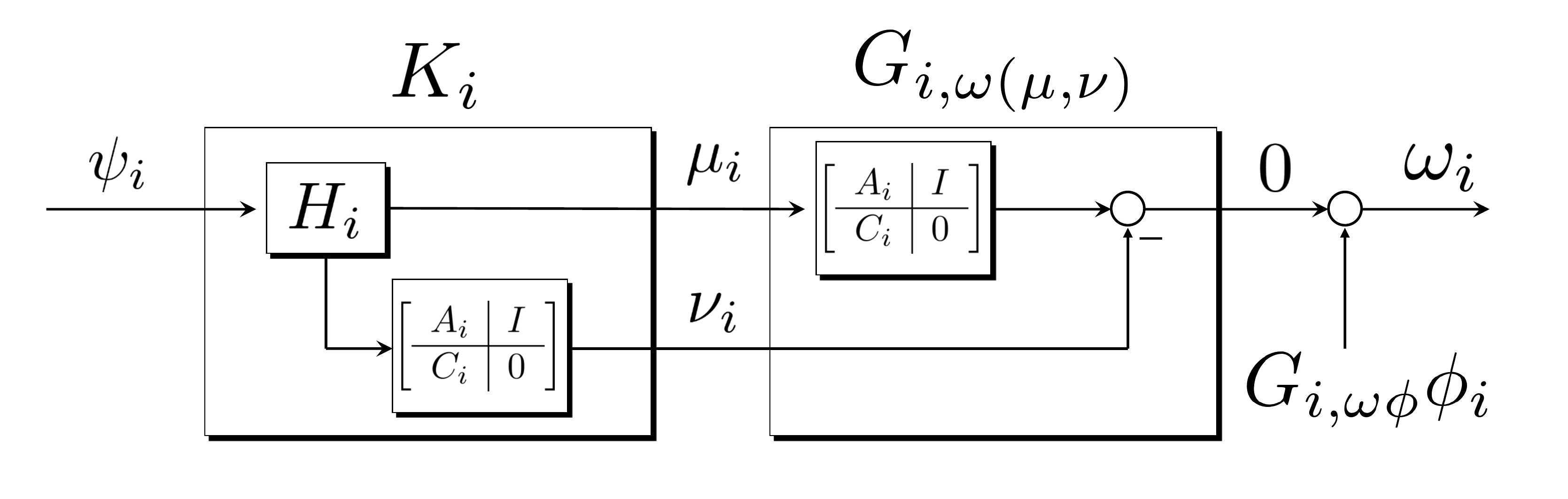}
\caption{Block diagram of $G_{i,\omega (\mu,\nu)}K_i$ and its internal structure where $G_{i,\omega \phi}:=W_i(sI-A_i)^{-1}L_i+Z_i$.}
\label{fig:block_GK}
\end{center}
\end{figure}

Fig.~\ref{fig:block_GK} depicts the block diagram of $G_{i,\omega (\mu,\nu)}K$ and its internal structure where $G_{i,\omega \phi}:=W_i(sI-A_i)^{-1}L_i+Z_i$.
The effect of $\mu_i$ to $\omega_i$ is canceled out by $\nu_i$ and it results in $\omega_i = G_{i,\omega \phi}\sum_{j \in \mathcal{N}_i}M_{ij}\omega_i$ in which the transfer matrix is invariant under the control.
Owing to the invariance, stability of the entire interconnected system can be preserved.
Note that, because the effect of the control inputs to the internal state $e_i$ is not canceled out, the response of the state varies under the control.
Note also that there always exists an observer gain $H_i$ such that $A_i+H_iC_i$ is Hurwitz because the pair $(A_i,C_i)$ is detectable from Assumption~\ref{assum:res}.

\emph{Solution:} In summary, the structure of the proposed disconnection-aware distributed dynamical attack detector is given by
\begin{equation}\label{eq:ob_str}
 \mathcal{O}_i:
 \left\{
 \begin{array}{l}
 \dot{\hat{x}}_i = \textstyle{A_i\hat{x}_i+L_i\sum_{j \in \mathcal{N}_i}M_{ij}\hat{w}_j + B_i r_i + \mu_i}\\
 \hat{w}_i = \textstyle{W_i \hat{x}_i + Z_i\sum_{j \in \mathcal{N}_i}M_{ij}\hat{w}_j + U_ir_i + \nu_i}\\
 \hat{y}_i = \textstyle{C_i\hat{x}_i + E_i\sum_{j \in \mathcal{N}_i}M_{ij}\hat{w}_j + D_ir_i}\\
 K_i:
 \left\{
 \begin{array}{cl}
 \dot{\chi}_i & = A_i \chi_i + \mu_i\\
 \mu_i & = H_i(\hat{y}_i-y_i)\\
 \nu_i & = -W_i\chi_i
 \end{array}
 \right.
 \end{array}
 \right.
\end{equation}
with observer gains $H_i$ such that $A_i+H_iC_i$ is Hurwitz for any $i \in \{1,\ldots,N\}$,
which gives a solution to Problem~\ref{prob}.


\section{Numerical Example}
\label{sec:num}

In this section, we examine how the proposed attack detector works through a numerical example of a power distribution network system.
For security issues in power distribution network systems, see~\cite{Teixeira2015Voltage,Isozaki2016Detection}.

The infrastructure of the distribution network is shown in Fig.~\ref{fig:DN} where ${\rm v}'_0$ represents the substation bus voltage magnitude that is assumed to be a given constant 230 [V].
Lines are represented by a series impedance $Z_k=R_k+jX_k$ or $Z'_k=R'_k+jX'_k$ and each customer has a DG with an inverter.
For each customer, the received voltage level is ${\rm v}_k$ and the voltage level at the point of connection with the distribution line is ${\rm v}'_k$.
At the $k$th connecting point, $S_k=P_k+j(q_{{\rm g},k}-q_{{\rm c},k})$ and $S'_k=P'_k+jQ'_k$ denote the power supplies from the DG and the distribution line, respectively. 
We employ the LinDistFlow model developed in~\cite{Baran1989Optimal} for power flow and voltage drop equations represented by
\[
 \left\{
 \begin{array}{cl}
 S'_k & = S'_{k+1}-S_k\\
 {{\rm v}'}^{2}_{k+1}&={{\rm v}'}_k^{2}-2\beta'_{k+1}(S'_{k+1})\\
 {\rm v}_k^2&={{\rm v}'}^{2}_k+2\beta_k(S_k)
 \end{array}\right.
\]
where $\beta_k(S_k) := R_kP_k+X_kQ_k$ and $\beta'_k(S'_k):=R'_kP'_k+X'_kQ'_k$.
The operation of the $k$th inverter is to regulate the voltage magnitude ${\rm v}_k$ by generating the reactive power $q_{{\rm g},k}$.
As the inverter dynamics, we employ the first-order model from the deviation of squared voltages between the reference value and the actual measured value to the generated reactive power, which is represented by
\[
 \dot{q}_{{\rm g},k} = -\frac{1}{\tau_k}q_{{\rm g},k}+\frac{1}{\tau_k}\kappa_k(\overline{{\rm v}}^2_k-{\rm v}_k^2)
\]
where $\tau_k$ is the time-constant of the $k$th inverter's response and $\kappa_k$ is the droop gain as in~\cite{Michelle2019Local}.
As a specific distribution network, we consider a benchmark European low-voltage distribution network in~\cite{Kai2014Benchmark} as in~\cite{Michelle2019Local}.
Although the detail is omitted, the distribution network system can be regarded as an interconnected system by defining the subsystem $\Sigma_k$ as each branch and the DG with an inverter and the interconnected signals $v_i$ and $w_i$ as the squared voltage, active power, and reactive power.
Moreover, it can numerically be confirmed that the interconnected system satisfies Assumption~\ref{assum:res}.

We now suppose that an attack is injected into the power distribution network system.
The $k_0$th customer's DG is assumed to be attacked and the squared voltage reference signal $\overline{{\rm v}}^2_{k_0}$ is fabricated.
As stated in Sec.~\ref{sec:uni}, we consider black-box attacks and the attack signal is modeled as a simple step function beginning at $t_{\rm a}$.
Accordingly, the fabricated squared reference voltage signal $\overline{{\rm v}}^2_{{\rm f},k_0}$ is represented by $\overline{{\rm v}}^2_{{\rm f},k_0}(t)=\overline{\rm v}^2_{k_0}(t)+\overline{a}\iota_{\{t:t \geq t_{\rm a}\}}(t)$ where $\overline{a}$ is a scaler value that represents the amplitude of the attack and $\iota_{\Omega}$ denotes the indicator function of the set $\Omega$.

For monitoring process, we design the proposed distributed attack detector.
Let $\Sigma_i$ be the subsystem composed of the $i$th customer and branch for $i\in \{1,\ldots,5\}$.
The internal state and the interconnection signal of $\Sigma_i$ are $q_{{\rm g},i}$ and $({{\rm v}'}^2_i,S'_i)$, respectively.
We suppose that each $q_{{\rm g},i}$ is measurable at each subobserver $\mathcal{O}_i$, namely, $y_i=q_{{\rm g},i}$, for attack detection.
Finally, the distributed observer is designed based on the structure~\eqref{eq:ob_str} where observer gains $H_i$ are determined based on the state-feedback linear quadratic regulator design method under certain weights.

\begin{figure}[h]
\begin{center}
\includegraphics[width=.98\linewidth]{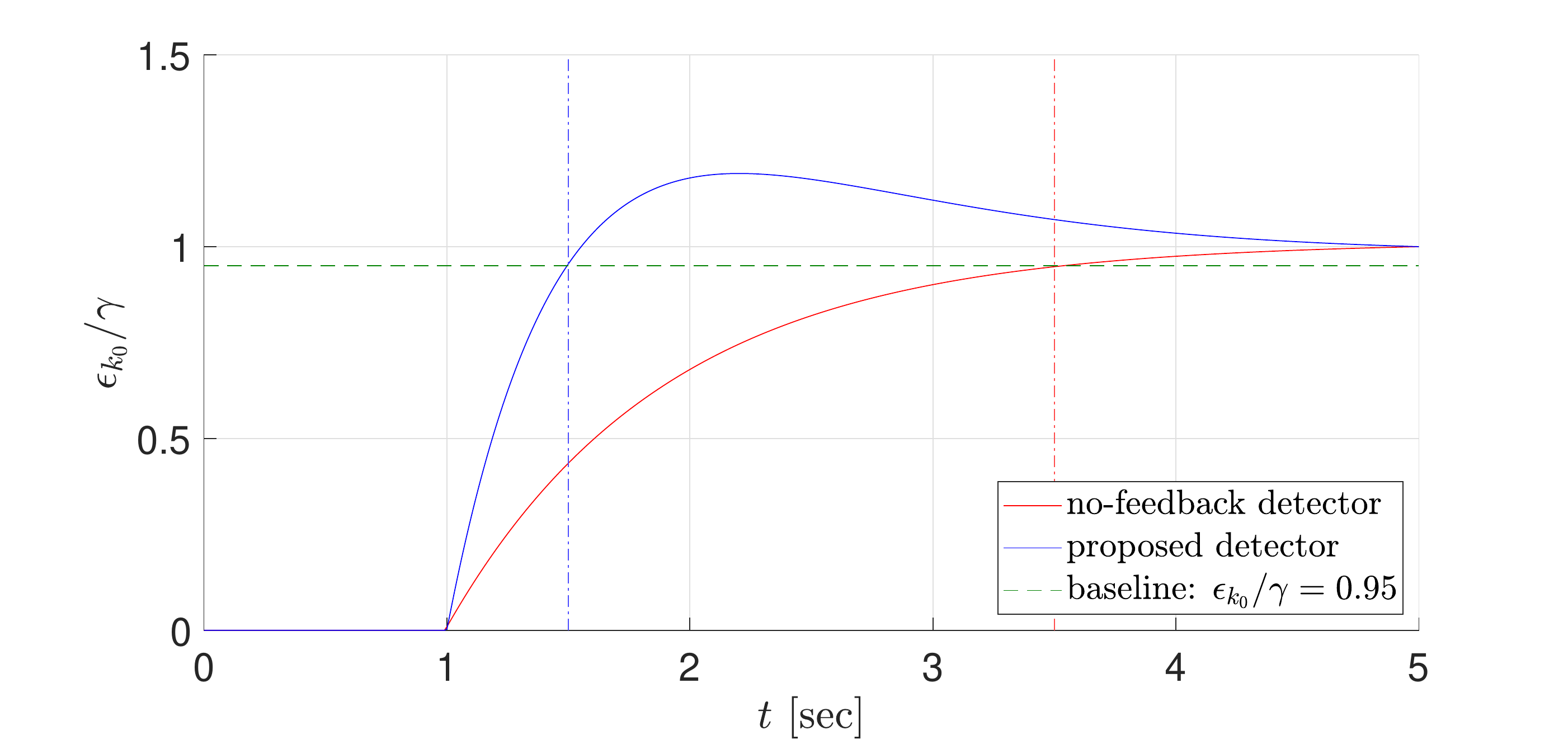}
\caption{Time series of normalized residuals $\epsilon_{k_0}/\gamma$ with the distributed attack detector without error feedback and the proposed distributed attack detector where $\overline{a}=1$, $t_{\rm a}=1$, and $\gamma$ is set to be the DC gain of the transfer function from the attack to the residual.}
\label{fig:resi_comp}
\end{center}
\end{figure}

Let us compare the time responses of residuals generated with the simple distributed attack detector without error feedback, namely,~\eqref{eq:obnomi} with $H_i=0$, and the proposed distributed attack detector.
Fig.~\ref{fig:resi_comp} shows the time response of the normalized residuals for the attack created by $\overline{a}=1$ and $t_{\rm a}=1$ with $k_0=4$,
where the normalized factor $\gamma$ is determined by the direct-current (DC) gain of the transfer function from the attack to the residual.
Let us suppose that 0.95 is taken as the baseline for the decision making, which is depicted by the broken line in Fig.~\ref{fig:resi_comp}.
When the value of the normalized residual exceeds the baseline, the decision of attack injection is made.
Then the times of attack detection are $t=3.5$ and $t=1.5$, which are represented by dash-dot lines, with the no-feedback detector and the proposed detector, respectively.
It can be observed that the detection time is much earlier owing to the estimation error feedback.

\begin{figure}[h]
  \begin{minipage}[t]{\linewidth}
    \centering
    \includegraphics[width=.98\linewidth]{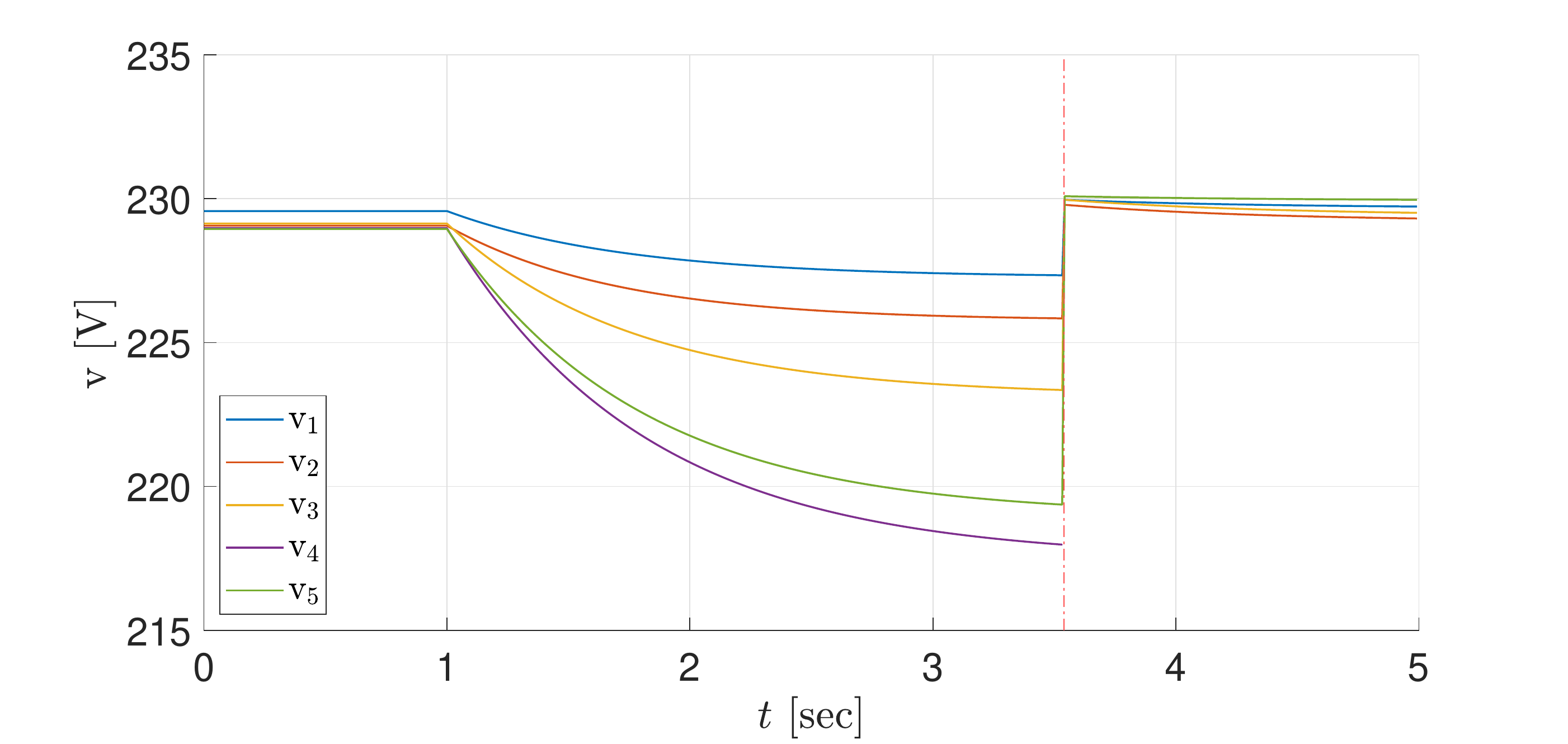}
    \subcaption{No-feedback distributed detector}
  \end{minipage}\\
  \begin{minipage}[t]{\linewidth}
    \centering
    \includegraphics[width=.98\linewidth]{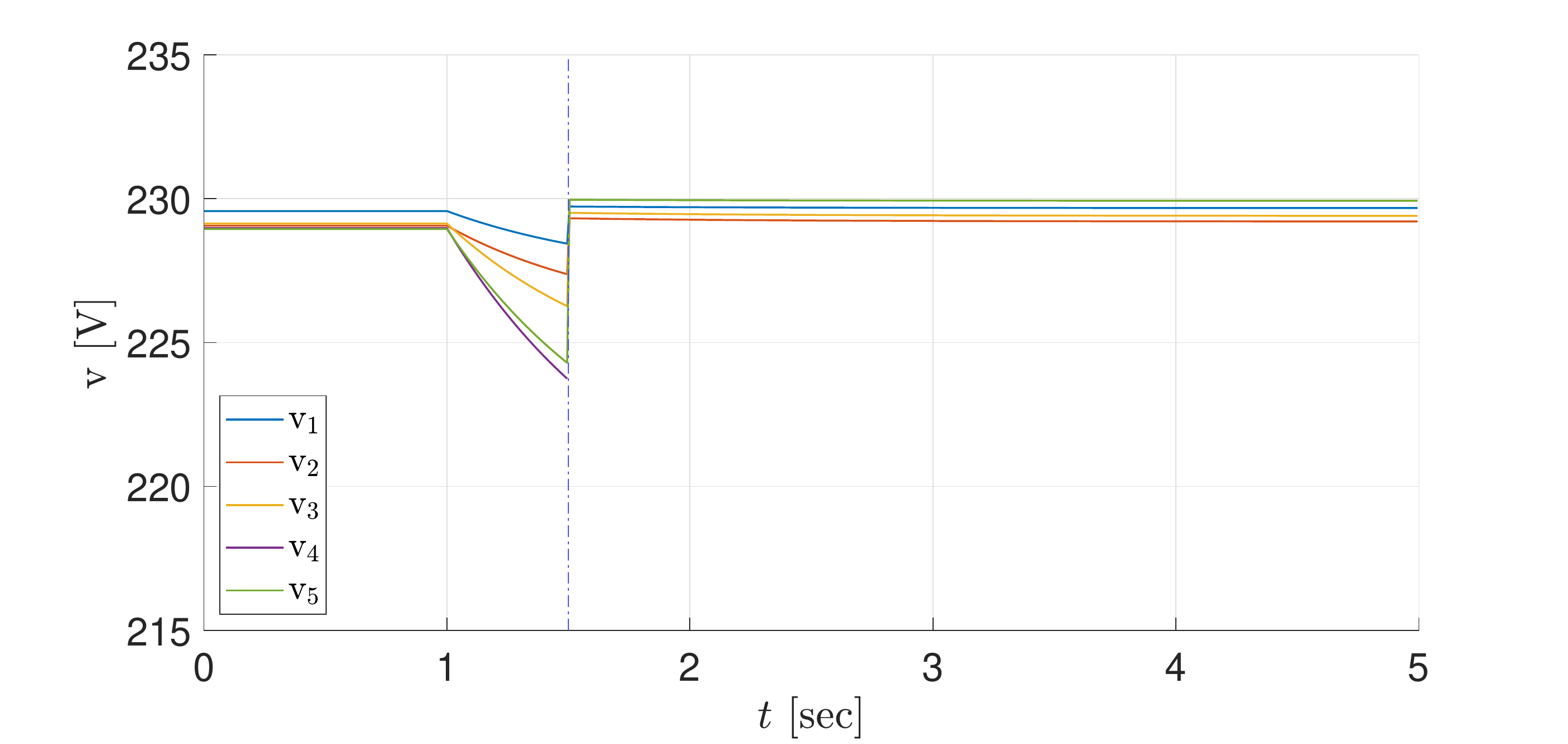}
    \subcaption{Proposed distributed detector}
  \end{minipage}
  \caption{Time series of each customer's voltage under the attack injected into the $k_0$th subsystem from $t_{\rm a}=1$ where $\Sigma_i$ is disconnected when the attack is detected at $\mathcal{O}_i$.}
  \label{fig:vol_det}
\end{figure}

The time series of each customer's voltage are shown in Fig.~\ref{fig:vol_det} where (a) is obtained with the no-feedback distributed detector and (b) is obtained with the proposed distributed detector.
In this example, the attack is made with the parameters $\overline{a}=-5\times 10^4$ and $t_{\rm a}=1$ with $k_0=4$.
The decision on disconnection in terms of $\Sigma_i$ is made when $|\overline{a}|\epsilon_{i}(t)/\gamma > 0.95$.
As shown in Fig.~\ref{fig:vol_det}, owing to the early detection achieved by the proposed detector the voltage deviations with respect to non-attacked customers are suppressed compared with the naive detector.

\begin{figure}[h]
\centering
\includegraphics[width=.98\linewidth]{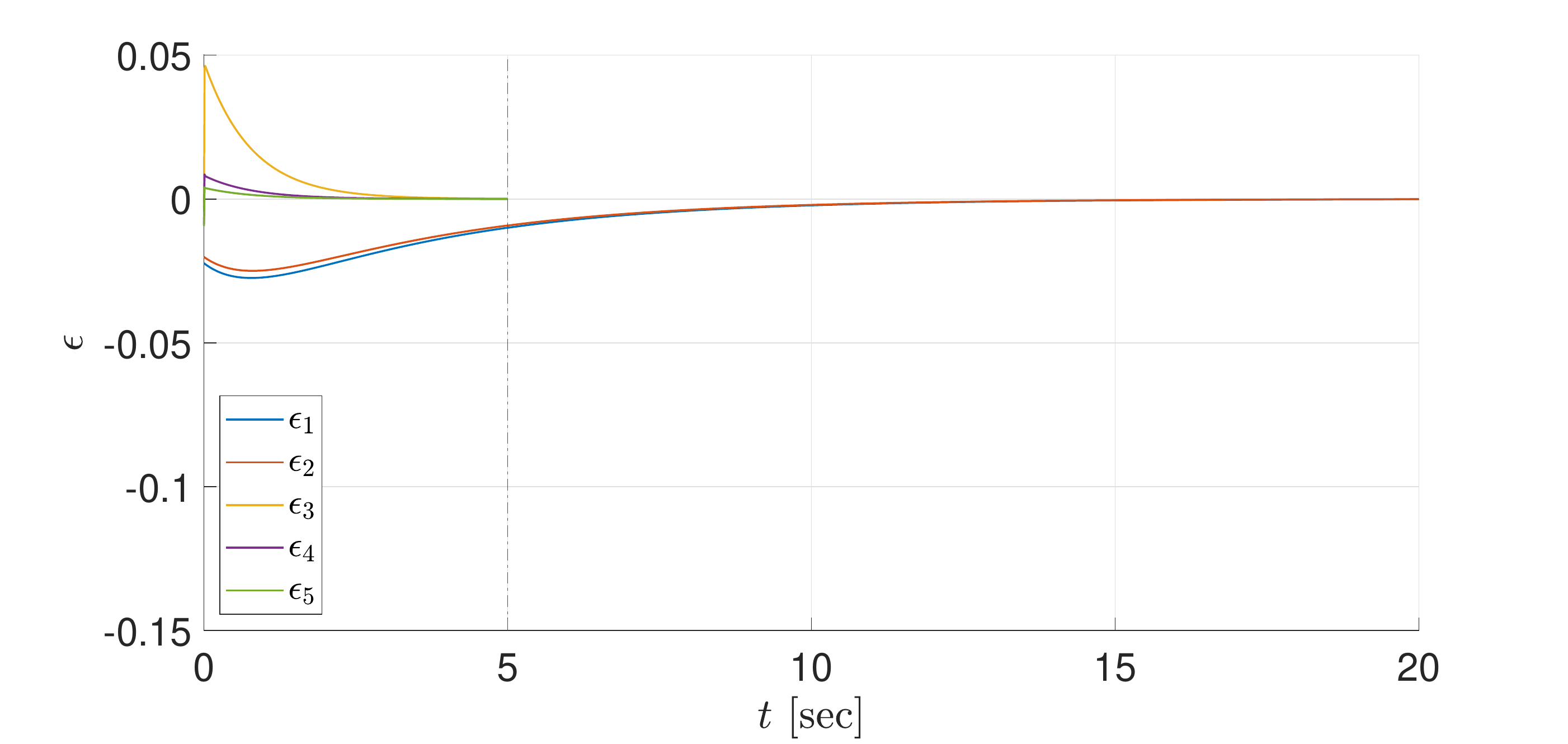}
\caption{Time series of the residual with the proposed distributed observer under the disconnection.}
\label{fig:res_prop}
\end{figure}

Finally, we confirm that the stability of the error dynamics is guaranteed under disconnection in the same setting of Fig.~\ref{fig:example}.
Fig.~\ref{fig:res_prop} illustrates the time series of the residual with the proposed attack detector where the observer gains are the same as the ones used in Fig.~\ref{fig:example}.
It can be observed that the stability is preserved even after the disconnection owing to the proposed structure.

\section{Conclusion}
\label{sec:conc}

In this paper, we have considered detection process of defense mechanism in physical layers towards attack-resilient cyber-physical systems.
In particular, we have paid attention to the disconnection-aware attack detector design problem and proposed a novel design method based on retrofit control.
As illustrated in the numerical example, the proposed attack detector enables early detection of attacks while maintaining the error dynamics to be stable even under network topology change caused by disconnection.

The important direction of future works is to develop an entire framework for the detection stage, including the decision step, the disconnection step, and the operation and restoration step.
For instance, while in the numerical example we employ a simple criterion for making a decision with a fixed baseline,
the criterion is desirable to be dependent on the time instant, the system model, the supposed attack, and also the variable network topology.
Moreover, worst-case analysis against white-box attacks is considered to be another direction of future works.

\bibliography{sshrrefs}

\begin{thebibliography}{28}
\providecommand{\natexlab}[1]{#1}
\providecommand{\url}[1]{\texttt{#1}}
\providecommand{\urlprefix}{URL }
\expandafter\ifx\csname urlstyle\endcsname\relax
  \providecommand{\doi}[1]{doi:\discretionary{}{}{}#1}\else
  \providecommand{\doi}{doi:\discretionary{}{}{}\begingroup
  \urlstyle{rm}\Url}\fi

\bibitem[{{Akhavan} et~al.(2019){Akhavan}, {Mohammadi}, {Vasquez}, and
  {Guerrero}}]{Akhavan2019Passivity}
{Akhavan}, A., {Mohammadi}, H.R., {Vasquez}, J., and {Guerrero}, J.M. (2019).
\newblock Passivity-based design of plug-and-play current-controlled
  grid-connected inverters.
\newblock \emph{IEEE Trans. Power Electron}.
\newblock (Early Access).

\bibitem[{Anderson et~al.(2019)Anderson, Doyle, Low, and
  Matni}]{Anderson2019System}
Anderson, J., Doyle, J., Low, S., and Matni, N. (2019).
\newblock System level synthesis.
\newblock \emph{Annual Reviews in Control}, 47, 364--393.

\bibitem[{{Baran} and {Wu}(1989)}]{Baran1989Optimal}
{Baran}, M. and {Wu}, F.F. (1989).
\newblock Optimal sizing of capacitors placed on a radial distribution system.
\newblock \emph{IEEE Trans. Power Del.}, 4(1), 735--743.

\bibitem[{Brogliato et~al.(2006)Brogliato, Lozano, Maschke, and
  Egeland}]{Bernard2006Dissipative}
Brogliato, B., Lozano, R., Maschke, B., and Egeland, O. (2006).
\newblock \emph{Dissipatie Systems Analysis and Control: Theory and
  Applications}.
\newblock Communications and Control Engineering. Springer, 2nd edition.

\bibitem[{C\'{a}rdenas et~al.(2011)C\'{a}rdenas, Amin, Lin, Huang, Huang, and
  Sastry}]{Alvaro2011Attacks}
C\'{a}rdenas, A.A., Amin, S., Lin, Z.S., Huang, Y.L., Huang, C.Y., and Sastry,
  S. (2011).
\newblock Attacks against process control systems: {R}isk assessment,
  detection, and response.
\newblock In \emph{Proc. the 6th ACM ASIA Conference on Computer and
  Communications Security}.

\bibitem[{Chakraborty et~al.(2018)Chakraborty, Alam, Dey, Chattopadhyay, and
  Mukhopadhyay}]{Anirban2018Adversarial}
Chakraborty, A., Alam, M., Dey, V., Chattopadhyay, A., and Mukhopadhyay, D.
  (2018).
\newblock Adversarial attacks and defences: {A} survey.
\newblock [Online]. Available: \url{https://arxiv.org/abs/1810.00069}.

\bibitem[{{Chong} et~al.(2019){Chong}, Umsonst, and
  Sandberg}]{Michelle2019Local}
{Chong}, M.S., Umsonst, D., and Sandberg, H. (2019).
\newblock Local voltage control of an inverter-based power distribution network
  with a class of slope-restricted droop controllers.
\newblock In \emph{Proc. the 8th IFAC Workshop on Distributed Estimation and
  Control in Networked Systems}.

\bibitem[{Dibaji et~al.(2019)Dibaji, Pirani, Flamholz, Annaswamy, Johansson,
  and Chakrabortty}]{Seyed2019Systems}
Dibaji, S.M., Pirani, M., Flamholz, D.B., Annaswamy, A.M., Johansson, K.H., and
  Chakrabortty, A. (2019).
\newblock A systems and control perspective of {CPS} security.
\newblock \emph{Annual Reviews in Control}, 47, 394--411.

\bibitem[{Ding et~al.(2018)Ding, Han, Xiang, Ge, and Zhang}]{Ding2018A}
Ding, D., Han, Q.L., Xiang, Y., Ge, X., and Zhang, X.M. (2018).
\newblock A survey on security control and attack detection for industrial
  cyber-physical systems.
\newblock \emph{Neurocomputing}, 275, 1674--1683.

\bibitem[{Ding(2013)}]{Ding2013Model}
Ding, S.X. (2013).
\newblock \emph{Model-Based Fault Diagnosis Techniques Design Schemes,
  Algorithms and Tools}.
\newblock Advances in Industrial Control. Springer, 2nd edition.

\bibitem[{Falliere et~al.(2011)Falliere, Murchu, and
  Chien}]{Nicolas2011Stuxnet}
Falliere, N., Murchu, L.O., and Chien, E. (2011).
\newblock W32. {Stuxnet Dossier}.
\newblock Technical report, Symantec.
\newblock [ONLINE] Available:
  \url{https:{\slash}{\slash}www.symantec.com{\slash}content{\slash}en{\slash}us{\slash}enterprise{\slash}media{\slash}security\_response{\slash}whitepapers{\slash}w32\_stuxnet\_dossier.pdf}.

\bibitem[{Ishizaki et~al.(2018)Ishizaki, Sadamoto, Imura, Sandberg, and
  Johansson}]{Ishizaki2018Retrofit}
Ishizaki, T., Sadamoto, T., Imura, J., Sandberg, H., and Johansson, K.H.
  (2018).
\newblock Retrofit control: {L}ocalization of controller design and
  implementation.
\newblock \emph{Automatica}, 95, 336--346.

\bibitem[{Ishizaki et~al.(2019)Ishizaki, Sasahara, Inoue, Kawaguchi, and
  Imura}]{Ishizaki2019Modularity}
Ishizaki, T., Sasahara, H., Inoue, M., Kawaguchi, T., and Imura, J. (2019).
\newblock Modularity-in-design of dynamical network systems: {R}etrofit control
  approach.
\newblock [Online]. Available: \url{https://arxiv.org/abs/1902.01625}.

\bibitem[{{Isozaki} et~al.(2016){Isozaki}, {Yoshizawa}, {Fujimoto}, {Ishii},
  {Ono}, {Onoda}, and {Hayashi}}]{Isozaki2016Detection}
{Isozaki}, Y., {Yoshizawa}, S., {Fujimoto}, Y., {Ishii}, H., {Ono}, I.,
  {Onoda}, T., and {Hayashi}, Y. (2016).
\newblock Detection of cyber attacks against voltage control in distribution
  power grids with {PVs}.
\newblock \emph{IEEE Trans. Smart Grid}, 7(4), 1824--1835.

\bibitem[{Kulpers and Fabro(2006)}]{Kulpers2006Control}
Kulpers, D. and Fabro, M. (2006).
\newblock Control systems cyber security: {D}efense in depth strategies.
\newblock Technical report, Idaho National Laboratory.
\newblock [Online]. Available:
  \url{https:{\slash}{\slash}inldigitallibrary.inl.gov{\slash}sites{\slash}sti{\slash}sti{\slash}3375141.pdf}.

\bibitem[{{Kushner}(2013)}]{Kushner2013Real}
{Kushner}, D. (2013).
\newblock The real story of {S}tuxnet.
\newblock \emph{IEEE Spectrum}, 50(3), 48--53.

\bibitem[{Lee et~al.(2016)Lee, Assante, and Conway}]{Lee2016Analysis}
Lee, R.M., Assante, M.J., and Conway, T. (2016).
\newblock Analysis of the cyber attack on the {Ukrainian} power grid.
\newblock \emph{Electricity Information Sharing and Analysis Center (E-ISAC)}.
\newblock [ONLINE] Available:
  \url{https:{\slash}{\slash}ics.sans.org{\slash}media{\slash}E-ISAC\_SANS\_Ukraine\_DUC\_5.pdf}.

\bibitem[{Mehra and Peschon(1971)}]{Mehra1971An}
Mehra, R.K. and Peschon, J. (1971).
\newblock An innovations approach to fault detection and diagnosis in dynamic
  systems.
\newblock \emph{Automatica}, 7(5), 637--640.

\bibitem[{Miller and Valasek(2015)}]{Charlie2015Remote}
Miller, C. and Valasek, C. (2015).
\newblock Remote exploitation of an unaltered passenger vehicle.
\newblock [ONLINE] Available:
  \url{http:{\slash}{\slash}illmatics.com{\slash}Remote\%20Car\%20Hacking.pdf}.

\bibitem[{{Pasqualetti} et~al.(2015){Pasqualetti}, {Dorfler}, and
  {Bullo}}]{Pasqualetti2015Control}
{Pasqualetti}, F., {Dorfler}, F., and {Bullo}, F. (2015).
\newblock Control-theoretic methods for cyberphysical security: {G}eometric
  principles for optimal cross-layer resilient control systems.
\newblock \emph{IEEE Control Systems Magazine}, 35(1), 110--127.

\bibitem[{Qu and Simaan(2014)}]{Zhihua2014Modularized}
Qu, Z. and Simaan, M.A. (2014).
\newblock Modularized design for cooperative control and plug-and-play
  operation of networked heterogeneous systems.
\newblock \emph{Automatica}, 50(9), 2405--2414.

\bibitem[{{Riverso} et~al.(2016){Riverso}, {Boem}, {Ferrari-Trecate}, and
  {Parisini}}]{Riverso2016Plug}
{Riverso}, S., {Boem}, F., {Ferrari-Trecate}, G., and {Parisini}, T. (2016).
\newblock Plug-and-play fault detection and control-reconfiguration for a class
  of nonlinear large-scale constrained systems.
\newblock \emph{IEEE Trans. Autom. Control}, 61(12), 3963--3978.

\bibitem[{{Sasaki} et~al.(2015){Sasaki}, {Sawada}, {Shin}, and
  {Hosokawa}}]{Sasaki2015Model}
{Sasaki}, T., {Sawada}, K., {Shin}, S., and {Hosokawa}, S. (2015).
\newblock Model based fallback control for networked control system via
  switched {Lyapunov} function.
\newblock In \emph{IECON 2015 - 41st Annual Conference of the IEEE Industrial
  Electronics Society}, 2000--2005.

\bibitem[{{Sasaki} et~al.(2017){Sasaki}, {Sawada}, {Shin}, and
  {Hosokawa}}]{Sasaki2017Model}
{Sasaki}, T., {Sawada}, K., {Shin}, S., and {Hosokawa}, S. (2017).
\newblock Model based fallback control for networked control system via
  switched {Lyapunov} function.
\newblock \emph{IEICE Trans. Fundamentals}, E100-A(10), 2086--2094.

\bibitem[{Shepard et~al.(2012)Shepard, Bhatti, and Humphreys}]{Daniel2012Drone}
Shepard, D.P., Bhatti, J.A., and Humphreys, T.E. (2012).
\newblock Drone hack: {S}poofing attacks demonstration on a civilian unmanned
  aerial vehicle.
\newblock \emph{{GPS World}}, 23(8), 30--33.

\bibitem[{Strunz et~al.(2014)Strunz, Abbasi, and {Abbey, C., et
  al.}}]{Kai2014Benchmark}
Strunz, K., Abbasi, E., and {Abbey, C., et al.} (2014).
\newblock Benchmark systems for network integration of renewable and
  distributed energy resources.
\newblock \emph{CIGRE Task Force}.

\bibitem[{{Teixeira} et~al.(2015){Teixeira}, {Paridari}, {Sandberg}, and
  {Johansson}}]{Teixeira2015Voltage}
{Teixeira}, A., {Paridari}, K., {Sandberg}, H., and {Johansson}, K.H. (2015).
\newblock Voltage control for interconnected microgrids under adversarial
  actions.
\newblock In \emph{2015 IEEE 20th Conference on Emerging Technologies Factory
  Automation (ETFA)}, 1--8.

\bibitem[{{Willsky} and {Jones}(1976)}]{Willsky1976A}
{Willsky}, A. and {Jones}, H. (1976).
\newblock A generalized likelihood ratio approach to the detection and
  estimation of jumps in linear systems.
\newblock \emph{IEEE Trans. Autom. Control}, 21(1), 108--112.

\end{thebibliography}
                                                                                                    







\end{document}